\documentclass[12pt]{article}
\usepackage{amsmath,amssymb,amsfonts,textcomp,times}
\usepackage[T1]{fontenc}
\usepackage{graphicx,color}
\begin{document}
\thispagestyle{empty}
\setcounter{page}{0}

\noindent
\hrule
\vspace{0.3cm}
\noindent
{\fontfamily{cmss}\fontseries{bx}\fontsize{15}{0}\selectfont
Lethality and synthetic lethality in the genome-wide
metabolic network of \textit{Escherichia coli}\\}
\vspace{-0.2cm}
\hrule
\vspace{1.5cm}
\noindent
{\sf
Cheol-Min Ghim$^{*\ddag}$, Kwang-Il Goh$^*$, and Byungnam Kahng$^{*\dag}$\\}
\vspace{-0.3cm}

\noindent
$^*$School of Physics, Seoul National University NS50, Seoul 151-747, Korea\\
$^{\dag}$Program in Bioinformatics, Seoul National University NS50, Seoul 151-747, Korea\\
$^{\ddag}$Present address: Center for Theoretical Biological Physics, 
University of California,\\
\indent \hspace{-13pt}San Diego, La Jolla, CA 92093
\vspace{2cm}

\renewcommand{\baselinestretch}{1.1}
\normalsize
\renewcommand{\thefootnote}{\fnsymbol{footnote}}
\noindent 
{\sf\large ABSTRACT}\\

\noindent
Recent genomic analyses on the cellular metabolic network show that 
reaction flux across enzymes are diverse and exhibit power-law behavior
in its distribution. While one may guess that the reactions with larger
fluxes are more likely to be lethal under the blockade of its catalyzing 
gene products or gene knockouts, we find, by in silico flux analysis,
that the lethality rarely has correlations with the flux level owing to 
the widespread backup pathways innate in the genome-wide metabolism of 
\textit{Escherichia coli}. Lethal reactions, of which the deletion 
generates cascading failure of following reactions up to the biomass 
reaction, are identified in terms of the Boolean network scheme as well as 
the flux balance analysis. The avalanche size of a reaction, defined as
the number of subsequently blocked reactions after its removal, turns out 
to be a useful measure of lethality. As a means to elucidate phenotypic 
robustness to a single deletion, we investigate synthetic lethality in 
reaction level, where simultaneous deletion of a pair of nonlethal 
reactions leads to the failure of the biomass reaction. Synthetic lethals
identified via flux balance and Boolean scheme are consistently shown to
act in parallel pathways, working in such a way that the backup machinery
is compromised.

\newpage

\section{Introduction}
\noindent
Complex machinery of cellular metabolism occurring in a living organism
makes up a part of autocatalytic network of biochemical reaction pathways. 
The reaction network in itself constitutes an intricate web in such a way as 
sharing intermediates. Yet another dimension of complexity comes from the 
tight control of reactions by functional proteins which are again under 
transcriptional, translational control as well as degradative and  other 
inductive regulations, making even a single pathway analysis formidable 
task. Only recently, advances in high-throughput experiments 
and the computing power incorporating diverse data sets collected in genomic 
research make it possible to construct cellular networks of metabolism 
in genome-wide perspectives.
At the same time, many quantitative theoretical methods including graph 
theories and other mathematical tools developed from diverse disciplines 
attract much attention to tackle the large-scale networks 
(Xia et al., 2004; Barab\'asi \& Oltvai, 2004). 
\par
In the early graph-theoretic approaches to the metabolic network, 
attention has been paid to the so-called scale-free feature of topological 
structure (Jeong et al., 2000), small-world-ness (Wagner and Fell, 2001), 
modularity (Girvan and Newmann, 2002) and hierarchical organization
(Ravasz et al., 2002). 
Despite the immanent specificity in cellular functions of various organisms, 
the connectivity, or number of connections each node (metabolite or associated 
reactions) has, is generally far from homogeneous.
In particular, this connectivity distribution of the metabolic network, 
as shared by many naturally occurring complex networks, follows a power law, 
meaning large deviations in spite of well defined average value. 
It is this context that borrows the term \textit{scale-free} network,
where hubs, nodes with large number of connections, play essential roles. 
When such hubs are removed or turned off, the whole system becomes 
vulnerable. Indeed, it was found (Jeong et al., 2001) that, 
for the yeast protein interaction network, hub proteins are more likely 
to be lethal than the others. 
\par
In the framework of networks, metabolic reactions and participating 
metabolites can be mapped into alternating nodes, where the outward(inward) 
connections from a reaction node indicate that those metabolites are 
produced(consumed) as a result of the reaction. Once constructing a directed 
bipartite graph in this way, we calculate graph-theoretic quantities that 
characterize the global topology and give a clue to assessing lethality
of metabolic reactions.
Then, we examine the correlations between the metabolic flux level and the 
lethality of each metabolic reaction using the flux balance analysis[FBA] 
(Edwards and P{\aa}lsson, 2000).
Here, by lethal, we mean the organism could rarely synthesize the 
indispensable biomass, or the flux of the biomass reaction is significantly 
reduced when that reaction is blocked or removed from the network, mimicking 
gene knockout experiments\footnote{More recently, Segr\`e et al. 
proposed alternative scheme phrased as ``minimization of metabolic 
adjustment[MOMA]'' for the phenotypic prediction of deletion mutants. 
Instead of assuming optimality in growth yield of deletion mutants, MOMA 
approximates metabolic phenotype by performing distance minimization in flux 
space, whereby the correlation with experimental results are improved. 
See Segr\`e 2002.}. One seemingly counterintuitive result is the absence of 
correlation between flux level and lethality (Fig.~2), which is related with 
the fact that the high-flux reactions have abundant bypasses or backup pathways. 
\par
We also invoke the Boolean network scheme, an idealization of the
metabolic network as a wiring of binary logic gates to elucidate the
pathway structure of the network on the logical basis. Considering the
knockout and consequent cascading failure in the metabolic reaction network
as an avalanche, we investigate the distribution of avalanche size defined as 
the number of reactions subsequently turned off on account of the removal 
of a target reaction to find it a good measure of lethality. 
The distribution follows a power law with the characteristic exponent 
around 2.5, pervasive throughout disparate model systems having 
self-organized criticality (Bak, 1996).
\par
In the latter part, we investigate the effects of simultaneous deletion of 
multiple reactions with a view to elucidating the interaction between them.
Though only a small portion of reactions lead to distinctive phenotype under 
their single deletion, it is nontrivial to make a prediction on how destructive
a dual deletion of nonlethal reactions. Actually, the effect can be aggravating 
or alleviating as well as simple sum of each, depending on their role played in 
the otherwise intact metabolic network. In particular, when a pair of nonlethal 
reactions are deleted to make no growth of cell, we call it 
synthetic lethal, metabolic homologue of the same term in genetics. 
Synthetic lethality in metabolic network is a manifestation of their 
complementary nature responsible for the buffering between alternative 
parallel pathways. We show the synthetic lethal pairs are distributed over
distinct the avalanche size of a pair of reactions 
is strongly correlated with its synthetic lethality.\\

\section{Materials and Methods}
\noindent
We use, with minor curation, the recent revision of \textit{in silico} 
model \textit{E. coli} (Reed et al., 2003), which was obtained by searching 
databases, such as LIGAND (http://www.genome.jp/kegg/ligand.html), 
EcoCyc (http://www.ecocyc.org), TC-DB (http://tcdb.ucsd.edu/), and referring 
to updated literatures on sequence annotation (Serres et al., 2001).
To mimic random or targeted mutation strains, a specific reaction is removed 
from the network and the resultant metabolic capabilities are to be assessed. 
For this purpose, we introduce a single pivotal reaction, the biomass 
reaction, originally formulated as a linear combination of essential metabolic 
reactions giving rise to the growth of the organism (Neidhardt and Umbarger, 1996).
Throughout the study, lethality of a certain reaction is determined by the 
flux of this biomass production, which is contingent to the ansatz of 
optimality that the selection pressure has imposed in the long history 
of evolution.\\

\subsection{Metabolic Network as a Graph}
\noindent
The overall map of metabolic reactions we study is a bipartite 
graph, composed of two different types of nodes, 627 metabolites and
their participating 1074 metabolic reactions including transport and 
exchange events. One type of nodes connect only to the other type 
of nodes in the networks\footnote{Depending upon the objectives, it can be 
projected to recover either of the single-mode metabolite network or 
reaction (enzyme) network.}. Each link between a pair of a metabolite 
and a reaction is directed, reflecting the metabolite is either 
consumed (substrate) or produced (product) or both in reactions. 
Of the 1074 reactions, 254 reversible reactions are decomposed into
two separate reactions catalyzed by the same enzyme, and possible 
multiple connections between metabolites via isozymes are taken as 
a single directed edge in the graph representation. 
627 distinct metabolites have either their intracellular or extracellular 
version or both, which sum into 761 distinct nodes of metabolites.
Once the metabolic network is reconstructed as a graph, we quantify, 
by various numerical measures, the lethality of each node in the 
wild-type strain and compare them with those of knockout mutants.\\

\subsection{Flux Balance Analysis (FBA)}
\noindent
For each metabolite in the metabolic reaction network, dynamic flux 
balance condition on the concentration $X_i$ can be written as 
\begin{equation} 
\label{eq1}
\frac{\partial X_i}{\partial t} =-\sum_{j=1}^n S_{ij}\phi_j~, 
\end{equation} 
where $\phi_j$ is the corresponding reaction rate (outward flux),
and $S_{ij}$ are stoichiometric coefficients involved in the 
metabolite $i\in\{1,2,\cdots,m\}$ participating in the reaction 
$j\in\{1,2,\cdots,n\}$ including transport and exchange reactions also. 
As an alternative to yet intractable kinetic models in genome-wide 
perspectives, mass conservation can be applied in the balanced state to 
give the stationary condition $\sum_{j}S_{ij}\phi_j=0$ for all $i$.
Stoichiometric matrix encodes the topology of the metabolic network 
and plays the role of weighted adjacency matrix in the corresponding graph.
Once the stoichiometric coefficients are given for all reactions,
we have, in general, under-determined situation($m<n$) where huge 
degeneracy in the null space of Eq.~(\ref{eq1}) is unavoidable.
Since, in reality, enzymatic capacity cannot be arbitrarily large, 
flux values are also constrained to vary in limited range, which can be
estimated via, say, the Michaelis-Menten type kinetics. 
Actually, this multiplicity in the feasible metabolic state can be 
considered a manifestation of the capability of a metabolic genotype. 
It makes physiological sense in that cells are expected to adapt 
themselves responding to different stresses and growth conditions.
Among those feasible metabolic phenotypes, FBA assumes the existence of
optimized point(s) of a certain objective function, called biomass reaction 
or growth flux, by utilizing further constraints based on the thermodynamic 
irreversibility of reactions.\\ 

\subsection{The Boolean Network}
\noindent
Metabolic reactions or genetic switches are seldom turned on or off. 
Instead, they can change by some fold, either up- or down-regulated, 
to make a physiological payoff. As a first approximation, Boolean 
idealization has long been introduced to reconstruct the genetic 
regulatory network as a directed graph (Kauffman, 1969) and was 
recently adopted to model the metabolic network (Lemke et al., 2004).
In Boolean reconstruction of the metabolic network, each node in the graph
is replaced with binary logic gate AND or OR. A metabolite would not cease 
to be existent in the network until all the reactions having that metabolite 
as a product are blocked, while a reaction does not take place any more 
with only a single absence out of substrate metabolites, which renders the 
metabolites Boolean disjunction, OR, and reactions conjunction, AND
as exemplified in Fig.~1. 
With all nodes given an initial condition, we iteratively relax 
the network till the network is settled in a fixed point.\\

\subsection{Cumulative Lethality Score}
\noindent
As a way to reveal the correlation between avalanche size and
lethality, we use the index, cumulative lethality score (CLS), the
original version of which was introduced in the context of protein
essentiality prediction (Jeong et al, 2003). Once a measure of lethality, 
say the flux level of each reaction, is proposed, we can make a
serial list of reactions assorted in descending order of the
proposed lethality measure. With the lethality criteria determined
through the viability under the single deletion of each reaction,
we can assign a binary lethality score, one or zero, into each
reaction depending on whether it is lethal or not. Summing up the
binary scores from the first rank to the $j$th rank in the value of the
proposed quantity, we get the CLS, say, $L(j)$. If the
proposed quantity is positively(negatively) correlated with the
lethality, \textit{i.e.} reactions holding high ranks are dominated 
by the binary lethality score one(zero), the normalized 
CLS manifests itself as a concave(convex) curve. Otherwise, if it 
is a random sequence of zero's and one's, $L(j)$ is given by a 
straight line. That is, the more correlated is the measure with 
the lethality, the higher curvature $L(j)$ develops.\\

\section{Results and Discussion}
\noindent
\subsection{Flux Level versus Lethality}
\noindent
Under aerobic condition with nine distinct carbon sources, we identify 
208 (19.6\%) common lethal reactions, the single blockade of which suffocates 
the biomass production\footnote{Throughout the analysis, acetate, 
alpha-ketoglutarate, glucose, glycerol, (L-,D-)lactate, L-malate, pyruvate, 
and succinate are used as the carbon sources. Though the lethality of 
a reaction does depend on which are used as carbon sources, 208 lethal 
reactions are shared by every sources and only 22 reactions show 
nutrient-dependent lethality.}. 
It is similar in fraction to the essential fraction of \textit{S. cerevisiae} 
genome, 18.7\% (Tong et al., 2001; Winzeler et al., 1999; Giaever et al., 2002).
\par 
The flux distribution in the wild-type metabolic network follows a 
power-law in Pareto's form, $P(\phi)\sim (\phi+\phi_0)^{-\alpha}$. 
Metabolic traffic is concentrated along a few `superhighway' reactions, 
while the vast majority of reactions are in charge of only a small 
flux (Almaas et al., 2004). 
In the meantime, inspired by the roles the main arteries, principal 
roads, or backbones  play in blood circulation, transportation, or 
data communications, we examine the possibility that the flux level 
should reflect the lethality of each reaction. Fig.~2 shows that 
the plausible correspondence between flux level and lethality is a 
mere conjecture to prove not true. 
In other words, there is no correlation at all between those quantities,
and high flux itself has nothing to do with the lethality of a reaction, 
which obviously contradicts our intuition. 
We also investigate the flux redistribution profile upon deletion of 
a high-flux reaction. Reaction fluxes are redistributed either locally 
or globally. Here, by local, we mean the case that very few reactions, 
having almost zero flux in the wild type, fully take over the flux of 
the reaction deleted. However, in the global redistribution, the flux of 
the removed reaction is shared over a large number of reactions to keep 
optimal biomass production.\\ 

\subsection{Avalanche Size versus Lethality}
\noindent
In the Boolean reconstruction of the \textit{E. coli} metabolic network,
41 lethal reactions are identified, which are all lethal in FBA also.
It is no wonder, if we consider that lethality in binary scheme
is minimal and more stringent than in the weighted version of FBA.
To quantify the effect of a single-node deletion, we define the avalanche
size of each reaction, in the Boolean scheme, as the number of reactions 
subsequently turned off on account of the removal of a target reaction. 
\par
As shown in the inset of Fig.~3, the avalanche size distribution for 
\textit{E. coli} metabolic network also displays a power law behavior, 
implying there exist a few reactions whose deletion triggers a large 
destructive avalanche cascade in metabolic reaction. In mathematical 
aspects, at criticality, there exists an infinite spanning cluster enough 
to reach the ultimate destination, biomass. Thus, avalanche size could be 
useful to find lethal reaction. Indeed, the reaction which generates a 
bigger avalanche size is more likely to be lethal as seen in the CLS plot 
(main frame of Fig.~3).  
\par
One may want to check the possibility that nearer nodes to the biomass 
reaction should be liable to be lethal. However, of the 374 distinct 
reactions that produce the substrates of biomass reaction, only 13 
of them are lethal in the Boolean scheme, making the identification of
lethal reactions nontrivial---proximity to the biomass reaction 
has nearly nothing to do with the lethality of a reaction.
\par
The notion of the avalanche in the network can be extended to the FBA 
scheme, where the avalanche size of a reaction is defined as the number 
of reactions whose flux levels under its knockout differ from those in 
the wild type. Actually, we used three different criteria of the 
avalanche: (i) reactions are newly turned on or off, 
(ii) flux level change exceeds an arbitrary cutoff value, and 
(iii) fractional changes in flux exceeds the cutoff value. There is,
however, little difference among the different counting schemes.
Fig.~4, drawn by using the criterion (i) above, reveals 
the absence of correlation between the flux level and the avalanche size 
both in the Boolean scheme and in FBA, which is consistent with the fact 
that flux level is irrelevant to assessing lethality.(Fig.~2)
\par
Due to the small-world-ness of the complex network, local perturbations 
are liable to propagate to the whole network leading to the sharing of 
load, which underlies the system-wide high flexibility. At the same
time, because of the scale-free-ness, avalanche cascade can either be 
long-ranged by making a large number of nodes bear parts of the `expenses' 
or be absorbed at a short distance from the source of perturbation, 
depending on the detailed functional characteristics of the reaction.
In other words, the response to single deletion perturbation of a single
reaction are too diverse to definitely predict how they would be, and so
they can be predicted in a probabilistic way (Kim et al., 2003). 
Identification of lethal reactions in metabolic network can be viewed in the
same footing. The effects of a node removal or a gene deletion are largely
negligible as a whole, which is manifested by the dominance of nonlethal
reactions. However, even if a reaction is not lethal, its potential damage 
to the network varies. It is this insufficiency in lethality assessing 
that raises the need for the knowledge of synthetic lethals 
in the next section.\\

\subsection{Synthetic Lethality in Metabolic Network}
\noindent
The close genetic relationships between genes which underlie the functional
buffering has been associated with the notion of synthetic lethality. It has 
been assessed in a high-throughput manner by the synthetic genetic array(SGA)
analysis (Tong et al., 2001).
Likewise, analysis of multiple-deletion mutants in genome-scale metabolic
network may shed light on novel topological features of backup pathways
leading to the robustness. In the restricted level of metabolism, such 
relationships can be revealed more clearly by performing the double reaction 
knockout experiments, which can be easily performed \textit{in silico}.  
Furthermore, such synthetic lethal pairs identified allow us to track the 
backup pathways and to visualize the precise genomic origin 
from which the metabolic phenotypic stability arises.
\par
When the glucose is used as a carbon source in aerobic condition, 
55 synthetic lethal pairs are identified out of all possible pairs of 
reactions(Table~I). Relatively small number of synthetic lethal pairs 
suggests high density of backup pathways for a given specific condition 
in the metabolic network and is in accordance with the case of the 
yeast (Papp et al, 2004). Among these, 33 (60\%) pairs are involved in 
the same subdivision of the reaction categories.  As expected, 
most of those homofunctional synthetic lethals usually work as the `simple' 
backup pathway: One of the pair is never used in the wild type but it 
almost completely takes over the flux of the blocked reaction, which amounts 
to 64\% of total synthetic lethal pairs, as depicted in Fig.~5(a). 
We note, however, the greater part of such simple takeover case might be
attributed to an artifact of the optimization scheme. Since there can 
be multiple solutions yielding the same maximized growth rate, and 
one cannot select a specific solution over the others without relevant 
physiological constraints, such as regulations in transcription level. 
Nevertheless, the effect of degeneracy on the (synthetic) lethality assay 
is not harmful since the deletion of lethal reaction unambiguously 
indicates no feasible solutions for nonzero biomass production. 
Interestingly, the homofunctional synthetic lethals of the other 
type, for which both the reactions are operational in the wild type, 
are mostly involved in the two subsystems of the pentose 
phosphate cycle and threonine-lysine metabolism. Excluding these particular
cases, 91\% of homofunctional synthetic lethals are simple, while, for the 
other 22 heterofunctional synthetic lethals, only 9 (41\%) of them are the 
simple backup pathways. In total, 25 pairs were both operational in the 
wild type and the remaining 30 are paired in one used and one unused in 
the wild type.
\par
We also investigate the synthetic lethality in the Boolean scheme, where
37 pairs of synthetic-lethal doublets are identified in addition to 41 
lethal singlets. Contrary to the lethal singlets which are dominated by 
cell envelope biosynthesis (78\%), synthetic lethals are scattered 
throughout diverse functional categories (Fig.~7).
\par
Conserved across both the network scheme is the proximity of the two 
reactions constituting a synthetic-lethal pair. Fig.~6 illustrates the 
distance distribution for the synthetic-lethal reaction pairs. 
As shown in the inset of Fig. 6, 94.5\% (97.3\%) of synthetic-lethal 
pairs in FBA(Boolean) scheme are apart in three steps and below, 
which stands out sharp contrast to the fraction 42.4\% for all pair of 
metabolites. If we eliminate, to make better sense of biochemical 
reaction pathways, currency metabolites or cofactors from the network, 
the fraction of synthetic-lethal pairs whose separation is three steps 
and below are 96.2\% and 91.1\% in FBA and Boolean scheme, 
respectively (main plot). Whether the analytical scheme is Boolean or 
FBA, long-ranged complicity of synthetic lethals 
through the intermediary of a `filamentary' single pathway like in 
Fig.~5(b) comprises only a small fraction($<$5\%) of synthetic-lethal 
pairs, and is rather an exception. 
\par 
Another important outcome regarding synthetic lethality is already shown 
in Fig.~3, where we measure the avalanche size of the synthetically 
lethal doublets and triplets in addition to that of the singlets. 
Synthetic lethal multiplets give rise to even higher correlation of 
their avalanche size with the lethality.
In effect, robustness in metabolic network stems from redundancy
in branched and parallel pathways. Conversely, lack of reaction pathways, 
whether it is due to the innate biochemical nature or to the incompleteness
of pathway database, lead to vulnerability. Hence, the more we know about 
a reaction pathway, the less probable it should contain lethal reactions.
In particular, we cannot completely rule out the latter possibility:
A quarter of the \textit{E. coli} genome is yet to be functionally 
assigned. Accumulated bias in molecular biology research, if any, 
might be considerable to our result that central pathway across the species, such as the 
citrate cycle or the glycolysis pathways, have very few lethal reactions 
(Fig.~7). 
However, at least for \textit{E. coli}, one of the best known 
bacteria yet studied, there are no good reasons to suspect such a bias. 
Moreover, our analytical results are compatible with the fact that a 
wide-spread strategy of antimicrobials is acting against cell wall 
synthesis(fosfomycin, cycloserine) or integrity(lysozyme).
Rather to be supposed is that the more important a reaction is, the 
better facilitated its backup pathways have been during evolution.\\

\section{Summary and Outlook}
\noindent
Systematic deletion study in a genome-wide view of model organisms help
reveal the organizing principles of the metabolic network and may shed 
light on how the selection has been embodied at the network levels, and 
especially the recent controversy surrounding the causes and evolution of 
the enzyme dispensability (Papp et al., 2004). 
As an index quantifying lethality in the graph-reconstructed metabolic 
reaction network, we propose the avalanche size of each reaction, the 
number of `dead' reactions due to the knockout of that reaction or its 
related gene products to show an even more remarkable interdependence 
than various measures yet proposed. By identifying synthetic lethals or
lethal multiplets in the genome-scale metabolic network under controlled 
environments, we see the emergence of phenotypic stability supported by 
rich backup pathways, which is shared by diverse realization of complex 
networks with scale-free nature.
Studies on multiple deletion mutations in metabolomic interaction network
can also be applied to natural metabolomic variations, reminiscent of 
single-nucleotide polymorphism, giving rise to practical buffering or 
phenotypic robustness under targeted mutations (Tong et al., 2001;
Tucker and Fields, 2003; Ooi et al., 2003). 
Furthermore, if we incorporate network analyses on metabolism with the 
gene-protein-reaction associations by including the other sectors of 
intra- and inter-cellular networks, it can be used in designing new 
microorganismal strains on the computer truly beyond the reductionist 
perspectives.\\

\section*{Note Added at Proof}
After completion of the present study, we learned of the work by 
Burgard et al. (2004), who introduced seemingly parallel concept of flux 
coupling analysis to our avalanche analysis in FBA scheme. Originally, 
the metabolic flux $\phi_1$ was defined as (directionally) 'coupled' to 
$\phi_2$ when a nonzero flux value $\phi_2$ implies a nonzero flux value 
$\phi_1$ in any stationary states of the system. Taking the contraposition, 
so as to interpret it in the context of the lethality, $\phi_1$ is coupled 
to $\phi_2$. Then the zero flux of $\phi_1$ implies the zero flux of $\phi_2$. 
While the avalanche size is correlated with the number of fluxes coupled with 
the biomass reaction, they are not mathematically identical, because the 
former refers only to the optimal flux distribution, whereas the latter 
covers all possible stationary states.

\section*{Acknowledgements}
\noindent
The authors thank J.~L.~Reed and B.~\O.~P{\aa}lsson for helpful comments. 
This work is supported by the KOSEF grants No.~R14-2002-059-01000-0 in the 
ABRL program and the MOST grant No.~M1~03B500000110.

\newpage
\section*{Appendices}
Abbreviation lists for metabolites and reactions (enzymes) in Fig.~1, 5 and Table I.
\begin{table*}[h]
\linespread 1
\tiny
\begin{center}
\begin{tabular}{llll}
\hline\hline
4r5au & 4-(1-D-Ribitylamino)-5-aminouracil & h & H$^+$ \\
adp & ADP & h2o & H$_2$O \\
akg & 2-Oxoglutarate & icit & Isocitrate \\
atp & ATP & mal-L & L-malate \\
cdp & CDP & nadh & Nicotinamide adenine dinucleotide, reduced \\
cit & Citrate & nadph & Nicotinamide adenine dinucleotide phosphate, reduced \\
co2 & CO$_2$ & oaa & Oxaloacetate \\
coa & Coenzyme A & pi & Phosphate \\
ctp & CTP & ppi & Diphosphate \\
db4p &3,4-dihydroxy-2-butanone 4-phosphate & q8 & Ubiquinone-8 \\
dcdp & dCDP & q8h2 & Ubiquinol-8 \\
dctp & dCTP & ribflv & Riboflavin \\
dmlz & 6,7-Dimethyl-8-(1-D-ribityl)lumazine & succ & Succinate \\
fad &  FAD & succoa & Succinyl-CoA \\
fadh2 & FADH2 & trdox & Oxidized thioredoxin \\
fmn & FMN & trdrd & educed thioredoxin \\
fum & Fumarate & & \\
\hline
ACONT  &  aconitase  & MTHFC  &  methenyltetrahydrofolate cyclohydrolase  \\
ADHEr  &  Acetaldehyde dehydrogenase  & MTHFD  &  methylenetetrahydrofolate dehydrogenase (NADP)  \\
ADK  &  adenylate kinase  & MTHFR2  & 5,10-methylenetetrahydrofolate reductase \\
ADK3  &  adentylate kinase (GTP)  & NADH6  &  NADH dehydrogenase (ubiquinone-8  \\
AGMT  &  agmatinase & NDPK  &  nucleoside-diphosphate kinase  \\
AKGDH  &   2-Oxogluterate dehydrogenase  & NDPK3  &  nucleoside-diphosphate kinase (ATP:CDP)  \\
ALAR  &  alanine racemase  & NDPK5  &  nucleoside-diphosphate kinase (ATP:dGDP)  \\
ALARi  &  alanine racemase (irreversible)  & NDPK7  &  nucleoside-diphosphate kinase (ATP:dCDP)  \\
ARGDC  &  arginine decarboxylase  & NDPK7  &  nucleoside-diphosphate kinase (ATP:dCDP) \\
ASNS  &  asparagine synthase (glutamine-hydrolysing)  & NDPK8  &  nucleoside-diphosphate kinase (ATP:dADP)  \\
ASNS2  &  asparagine synthetase  & O2t  &  O$_2$ transport (diffusion) \\
ASPTA  &  aspartate transaminase  & ORNDC  &  Ornithine Decarboxylase  \\
BIOMASS  &  Biomass reaction  & PDH  &  pyruvate dehydronase  \\
CBMK  &  Carbamate kinase  & PGK  &  phosphoglycerate kinase  \\
CBPS  &  carbamoyl-phosphate synthase (glutamine-hydrolysing)  & PIabc  &  phosphate transport via ABC system  \\
CO2t  &  CO$_2$ transporter via diffusion  & PIt2r  &  phosphate reversible transport via symport  \\
CS  &  citrate synthase & PPC  &  phosphoenolpyruvate carboxylase  \\
DB4PS  & 3,4-Dihydroxy-2-butanone-4-phosphate & PTAr  &  phosphotransacetylase  \\
DHORD2  &  dihydoorotic acid dehydrogenase (quinone8)  & RBFK  &  riboflavin kinase \\
DHORD5  &  dihydroorotic acid (menaquinone-8)  & RBFSa  &  riboflavin synthase \\
DKMPPD  & 2,3-diketo-5-methylthio-1-phosphopentane degradation reaction & RBFSb  &  riboflavin synthase \\
DKMPPD2  & 2,3-diketo-5-methylthio-1-phosphopentane degradation reaction & RNDR  &  ribonucleoside-diphosphate reductase (ADP)  \\
EX\underbar{\ }urea  &  trans-system-boundary (irreversible)  & RNDR2  &  ribonucleoside-diphosphate reductase (GDP)  \\
FMNAT  &  FMN adenylyltransferase  & RNDR3  &  ribonucleoside-diphosphate reductase (CDP)  \\
FRD2  &  fumarate reductase  & RNTR  &  ribonucleoside-triphosphate reductase (ATP)  \\
FUM  &  fumarase  & RNTR2  &  ribonucleoside-triphosphate reductase (GTP)  \\
GALU  &  UTP-glucose-1-phosphate uridylyltransferase  & RNTR3  &  ribonucleoside-triphosphate reductase (CTP)  \\
GALUi  &  UTP-glucose-1-phosphate uridylyltransferase (irreversible)  & RPE  &  ribulose 5-phosphate 3-epimerase  \\
GAPD  &   glyceraldehyde-3-phosphate dehydrogenase & SUCCt2b  &  Succinate efflux via proton symport  \\
GARFT  &  phosphoribosylglycinamide formyltransferase  & SUCD1i  &  succinate dehydrogenase  \\
GART  &  GAR transformylase-T  & SUCD4  &  succinate dehyrdogenase  \\
GHMT2  &  glycine hydroxymethyltransferase  & SUCOAS  &  succinyl-CoA synthetase (ADP-forming)  \\
GLUDy  &  glutamate dehydrogenase (NADP)  & TALA  &  transaldolase  \\
GLUSy  &  glutamate synthase (NADPH)  & THDPS  &  tetrahydropicolinate succinylase  \\
GND  &  phosphogluconate dehydrogenase  & THRAr  &  Threonine Aldolase  \\
HCO3E  &  HCO$_3$ equilibration reaction  & THRS  &  threonine synthase  \\
HSK  &  homoserine kinase  & TKT  &  transketolase  \\
HSST  &  homoserine O-succinyltransferase  & TKT2  &  transketolase  \\
ICDHyr  &  isocitrate dehydrogenase (NADP)  & TRPS  &  tryptophan synthase (indoleglycerol phosphate)  \\
ICL  &  Isocitrate lyase  & TRPS3  &  tryptophan synthase (indoleglycerol phosphate)  \\
KAS14  &  b-ketoacyl synthase  & UREAt  &  Urea transport via facilitate diffusion  \\
KAS15  &  b-ketoacyl synthase  & VALTA  &  valine transaminase  \\
MALS  &  malate synthase  & VPAMT  &  Valine-pyruvate aminotransferase  \\
MDH  &  malate dehydronase  &  &  \\
\hline\hline
\end{tabular}
\end{center}
\end{table*}

\newpage
\section*{References}
\begin{description}
\baselineskip 16pt
\item{Almaas, E., Kovacs, B., et al.}, 2004.
Global organization of metabolic fluxes in the bacterium 
\textit{Escherichia coli}. Nature 427, 839--843.

\item{Bak, P.,} 1996. How Nature Works: the Science of
Self-Organized Criticality. Copernicus, New York.

\item{Barab\'asi, A. L., Oltvai, Z. N.,} 2004. 
Network biology: understanding the cell's functional organization.
Nat. Rev. Genet. 5, 101--113.

\item{Burgard, A. P., Nikolaev, E. V., et al.,} 2004. 
Genome Res. 14, 301--312.

\item{Edwards, J.~S., P{\aa}lsson, B.~\O.,} 2000.
The \textit{Escherichia coli} MG1655 in silico metabolic genotype: 
Its definition, characteristics, and capabilities. 
Proc. Natl. Acad. Sci. USA 97, 5528--5533.

\item{Giaever, G., Chu, A.~M., et al.,} 2002. 
Functional profiling of the Saccharomyces cerevisiae genome.
Nature 418, 387--391.

\item{Girvan, M., Newman, M.~E.~J.,} 2002.
Community structure in social and biological networks.
Proc. Natl. Acad. Sci. USA 99, 7821--7826.

\item{Jeong, H., Tomber, B., et al.,} 2000. 
The large-scale organization of metabolic networks. Nature 407, 651--654.

\item{Jeong, H., Mason, S.~P., et al.,} 2001. 
Lethality and centrality in protein networks. Nature 411, 41--42.

\item{Jeong, H., Oltvai, Z.~N., Barab\'asi, A.-L.,} 2003.
Prediction of protein essentiality based on genomic data. 
Complexus 1, 19--28.

\item{Kauffman, S.~A.,} 1969. 
Metabolic stability and epigenesis in randomly constructed genetic nets.
J. Theor. B
iol. 22, 437--467.

\item{Kim, J.-H., Goh, K.-I., et al.,} 2003.
Probabilistic prediction in scale-free networks: Diameter changes.
Phys. Rev. Lett. 91, 058701-058704.

\item{Lemke, N., Her\'edia, F., et al.,} 2004.
Essentiality and damage in metabolic networks. 
Bioinformatics 20, 115-119.

\item{Neidhardt, F.~C., Umbarger, H.~E.,} 1996. 
\textit{Escherichia coli} and Salmonella. Am. Soc. Microbiol., 
Washington, DC, Vol.~1, pp.~13--16.

\item{Ooi, S.~L., Shoemaker, D.~D., Boeke, J.~D.,} 2003.
DNA helicase gene interaction network defined using synthetic lethality
analyzed by microarray. Nat. Genetics 35, 277--286.

\item{Papp, B., P\'al, C., Hurst, L.~D.,} 2004. 
Metabolic network analysis of the causes and evolution of enzyme
dispensability in yeast. Nature 429, 661--664 and references therein.

\item{Ravasz, E., Somera, A.~L., et al.,} 2002. 
Hierarchical organization of modularity in metabolic networks.
Science 297, 1551--1555.

\item{Reed, J.~L., Vo, T.~D., et al.,} 2003. 
An expanded genome-scale model of \textit{Escherichia 
coli} K-12 (iJR904 GSM/GPR). Genome Biol. 4, R54.

\item{Segre, D., Vitkup, D., et al.,} 2002. 
Analysis of optimality in natural and perturbed metabolic network. 
Proc. Nat. Acad. Sci. USA 99, 15112--15117.

\item{Serres, M.~H., Gopal S., et al.,} 2001.
A functional update of the \textit{Escherichia coli} K-12 genome.
Genome Biol. 2, research/0035.

\item{Tong, A.~H.~Y., Evangelista, M., et al.,} 2001. Systematic 
genetic analysis with ordered arrays of yeast deletion mutants.
Science 294, 2364-2368.

\item{Tucker, C.~L., Fields, S.,} 2003.
Lethal combinations. Nat. Genetics 35, 204--205.

\item{Wagner, A., Fell, D.~A.,} 2001.
The small world inside large metabolic networks.
Proc. R. Soc. Lond. B 268, 1803--1810.

\item{Winzeler, E.~A., Shoemaker, D.~D., et al.,} 1999. 
Functional characterization of the S. cerevisiae genome by gene deletion 
and parallel analysis. Science 285, 901--906.

\item{Xia, Y., Yu, H., et al.,} 2004. 
Analyzing cellular biochemistry in terms of molecular networks. 
Ann. Rev. of Biochem. 73, 1051--1087.
\end{description}

\newpage
\noindent{\sf\large Figure Legends\\}
\vspace{-10pt}

\noindent
{\sf Fig.~1\\}\noindent 
A subgraph of the citrate cycle in \textit{E. coli} metabolic network. In 
Boolean scheme, metabolites(ellipses) are treated as Boolean disjunction(OR), 
while the reactions(rectangles) as conjunction(AND). If this graph were 
isolated, though actually not the case, from the other 
reactions and metabolites, the metabolite \texttt{coa} would be no more 
supplied only when both the reactions \texttt{SUCOAS} and \texttt{CS} are 
blocked. On the other hand, the reaction \texttt{AKGDH} would not be 
operational when either of the metabolites \texttt{coa} or \texttt{akg} 
is knocked out. In this hypothetical subnetwork severed from the other 
part, bold red arrows indicate blocked reaction paths due to the knockout
of the reaction \texttt{CS}. Defining avalanche size as the number of 
reactions subsequently turned off on account of the removal of a target 
reaction, reaction CS has avalanche size as big as three.\\
\vspace{-10pt}

\noindent
{\sf Fig.~2\\}\noindent 
Normalized cumulative lethality score(CLS) with respect to the wild-type flux 
levels grown on distinct carbon sources, where 0(1) in the ordinate 
corresponds to the highest(lowest) ranker in the flux level. CLS against flux 
level shows no remarkable convexity, if any, implying they cannot be a 
lethality measure.\\

\noindent
{\sf Fig.~3\\}\noindent 
Normalized cumulative lethality scores drawn for the avalanche size of 
single and multiple deletion in the Boolean version of the \textit{E. coli}
metabolic network, where 0(1) in the ordinate corresponds to the highest(lowest) 
ranker in the flux level. Inset: Boolean avalanche size distribution under 
single targeted deletion of each reaction follows a power-law with the 
(noncumulative) exponent 2.5.\\
\vspace{-10pt}

\noindent
{\sf Fig.~4\\}\noindent 
Predicted lethal (red circles) and nonlethal (blue crosses) reactions in the 
avalanche size and the wild-type flux level determined by FBA (main frame) and the 
Boolean scheme (inset). In both the FBA and Boolean schemes, lethal reactions 
have larger avalanche size and are dispersed along wide range of flux level 
at the same time. Here, only the reactions being turned on in the wild-type 
strain are plotted.\\
\vspace{-10pt}

\noindent
{\sf Fig.~5\\}\noindent 
(a) As a consequence of NDPK7(nucleoside-diphosphate kinase, ATP:dCDP) 
removal, the reaction RNDR3(ribonucleoside-diphosphate reductase) is 
also blocked(dotted red lines), but RNTR3(ribonucleoside-triphosphate 
reductase) is newly turned on(solid (sky-)blue lines), 
with NDPK3(nucleoside-diphosphate kinase, ATP:CDP) flowing more flux 
to make up for the growth flux. Here, only three reactions are retuned, 
two of which depicted by filled blocks constitute a synthetic-lethal pair 
in both the Boolean and FBA schemes. The flux changes are shown in units 
of mm/g DW-hr.
(b) \textit{E. coli} can subsist without the reaction \texttt{SUCD4}, 
but, in the absence of \texttt{SUCD4}, all the reactions \texttt{DB4PS}, 
\texttt{RBFSa}, \texttt{RBFSb}, \texttt{RBFK}, and \texttt{FMNAT} become 
lethal. In particular, \texttt{DB4PS} has a `long-range correlation' with 
\texttt{SUCD4} along with the intermediary lethal reactions. Other parts 
of the network that the avalanche cascade does not reach are omitted 
for brevity.\\
\vspace{-10pt}

\noindent
{\sf Fig.~6\\}\noindent 
Synthetic lethal pairs tend to be closer neighbors than arbitrary pairs of 
reactions. Here, ALL refers to all pairs of reactions, while BOOL(FBA) 
stands for the Boolean(FBA) synthetic-lethal pairs. In the main histogram,
distance between a pair of metabolites is measured without counting the 
connection via currency metabolites or cofactors, while the inset is 
obtained by counting every conections. Reactions sharing either substrates
or products are regarded unit distance apart from each other. Across the 
analytical schemes, almost all ($\gtrsim$95\%) synthetic-lethal pairs are 
only three steps apart and below, in sharp contrast with the broad 
distribution of distances between arbitrary pairs. 
See the text for details.\\
\vspace{-10pt}

\noindent
{\sf Fig.~7\\}\noindent 
Classification of lethal and synthetic-lethal reactions according to the 
functional categories. Frequencies are normalized by total number of lethal 
singlets and doublets in both the Boolean and FBA scheme, respectively.\\
\vspace{-10pt}

\newpage
\noindent{\sf\large Table Legend\\}

\noindent
{\sf Table~1\\}\noindent 
List of synthetic-lethal reactions. Upper(lower) two cells include 
synthetic lethals belonging to the same(distinct) functional categories, 
and the wild-type flux of each reaction is given in units of mm/g DW-hr.\\

\newpage 
\begin{figure}[h!]
\centering
\includegraphics[height=7cm]{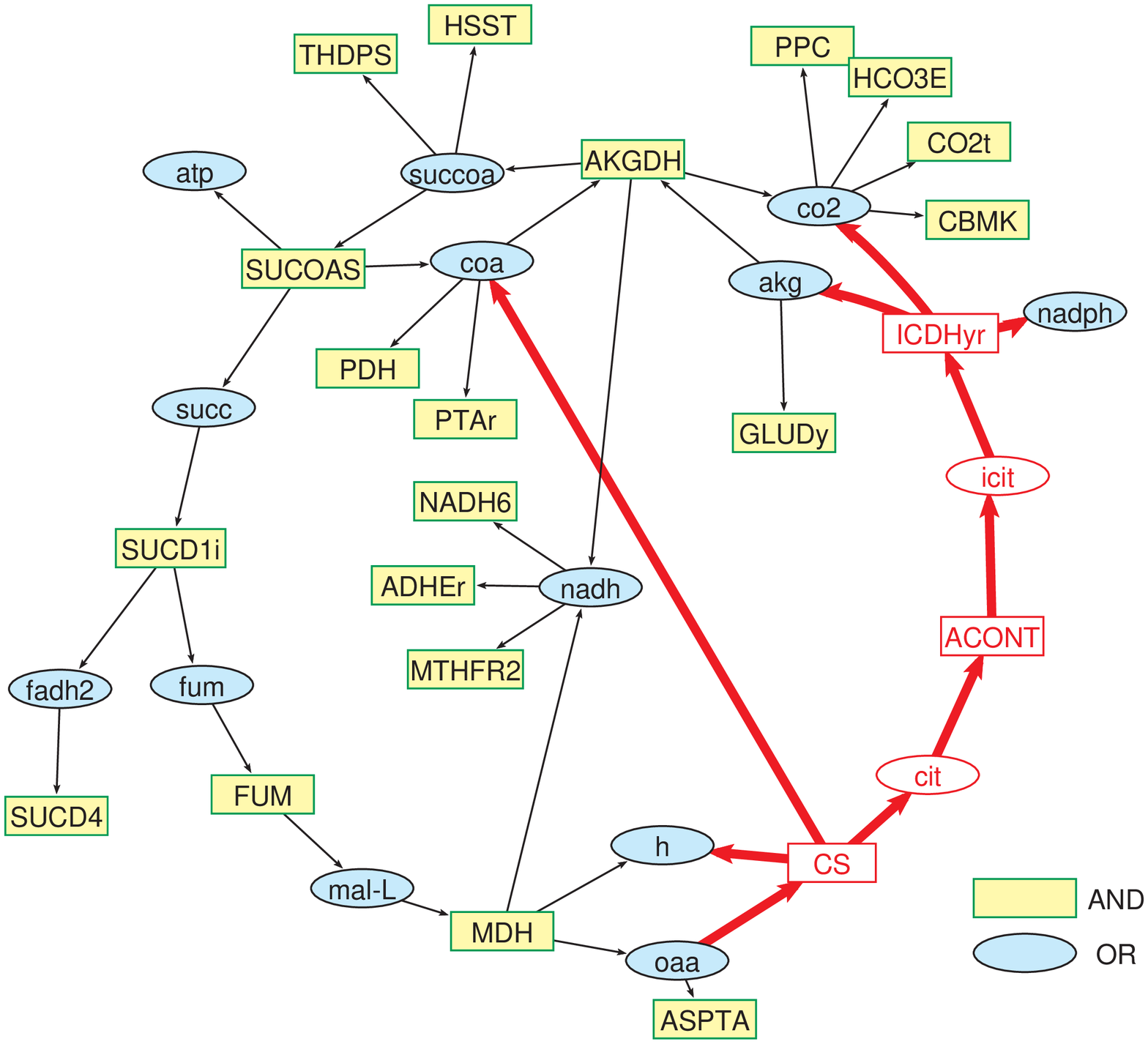}
\caption{Ghim, Goh \& Kahng}
\end{figure}

\newpage
\begin{figure}[h!]
\centering
\includegraphics[height=6cm]{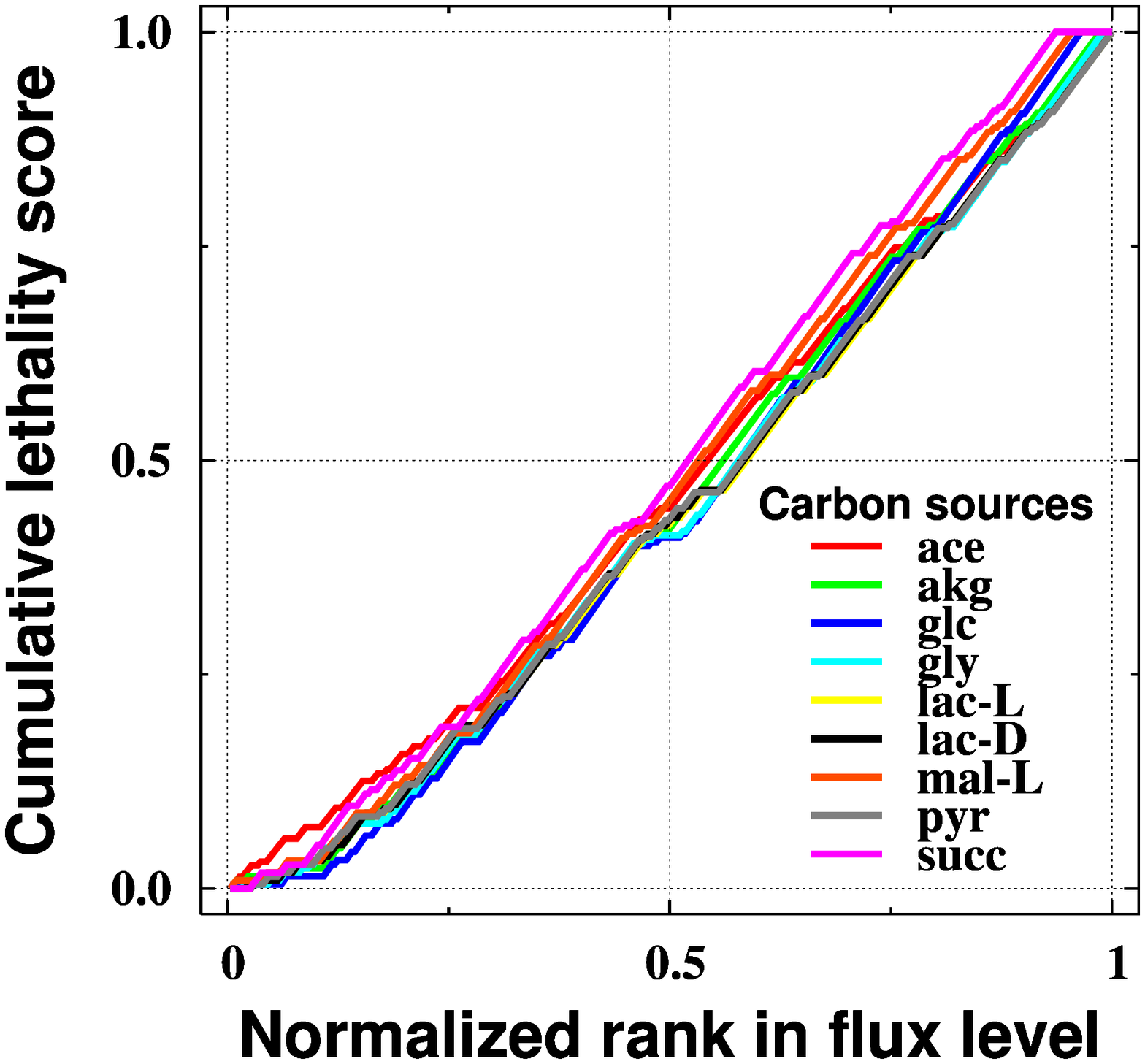}
\caption{Ghim, Goh \& Kahng}
\end{figure}

\newpage
\begin{figure}[h!]
\centering
\includegraphics[height=6cm]{FIG3.eps}
\caption{Ghim, Goh \& Kahng}
\end{figure}

\newpage
\begin{figure}[h!]
\centering
\includegraphics[height=6cm]{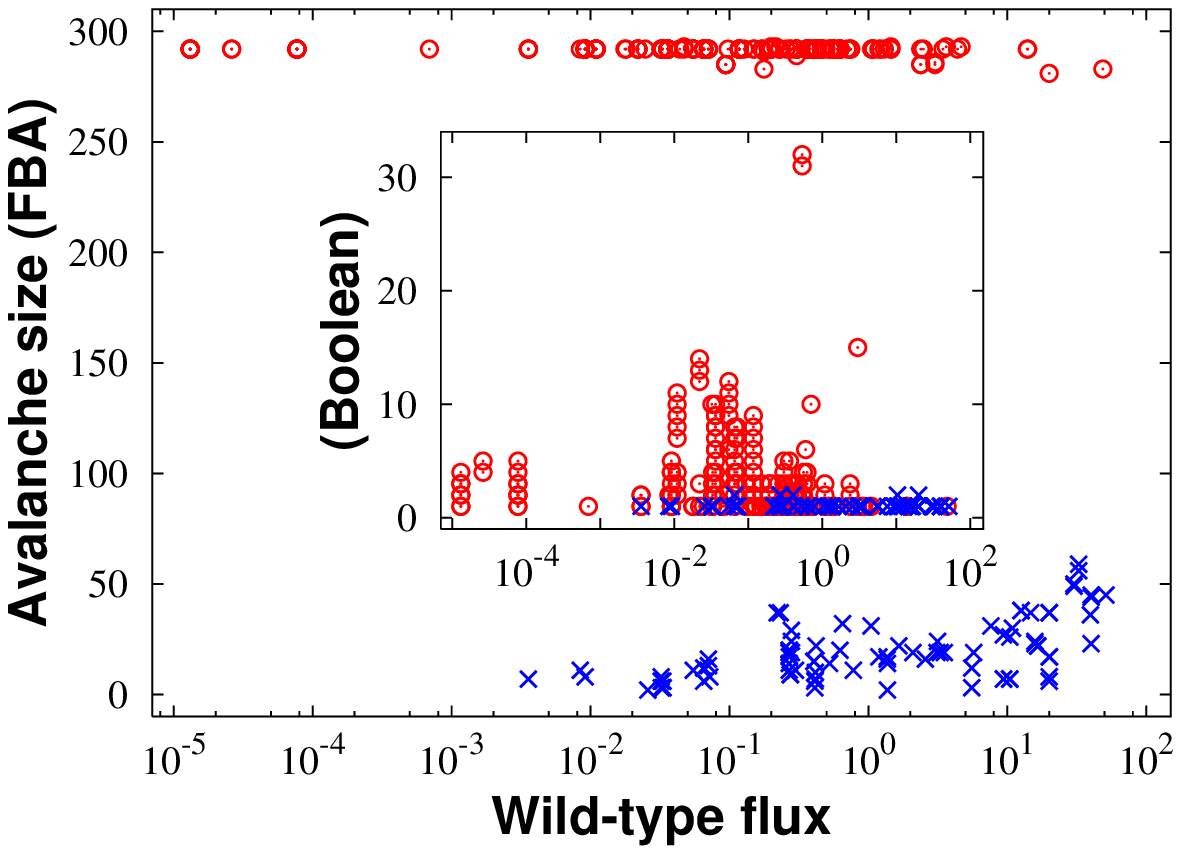}
\caption{Ghim, Goh \& Kahng}
\end{figure}

\newpage
\begin{figure}[h!]
\centering
\includegraphics[height=8cm]{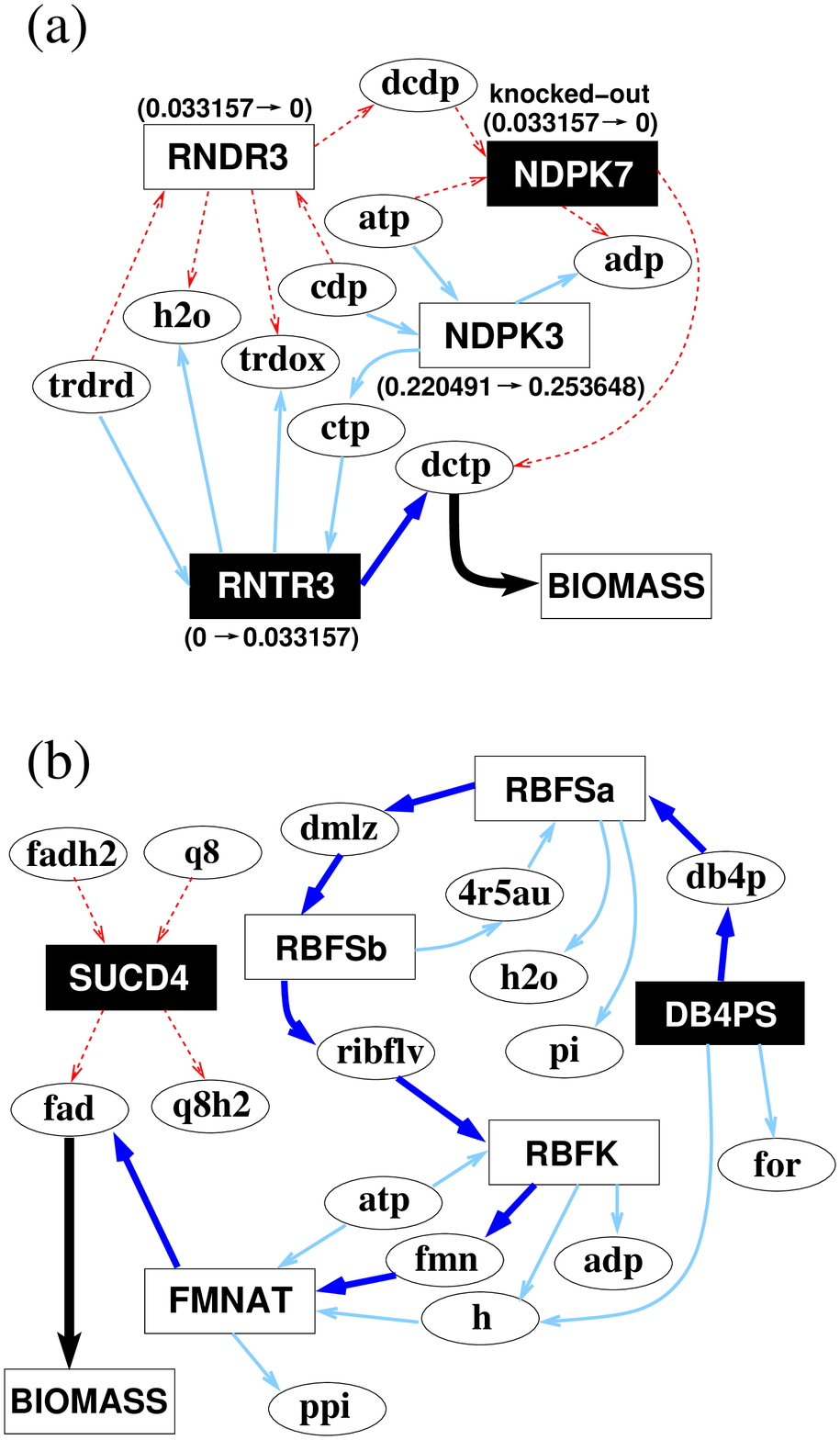}
\caption{Ghim, Goh \& Kahng}
\end{figure}

\newpage
\begin{figure}[h!]
\centering
\includegraphics[height=6cm]{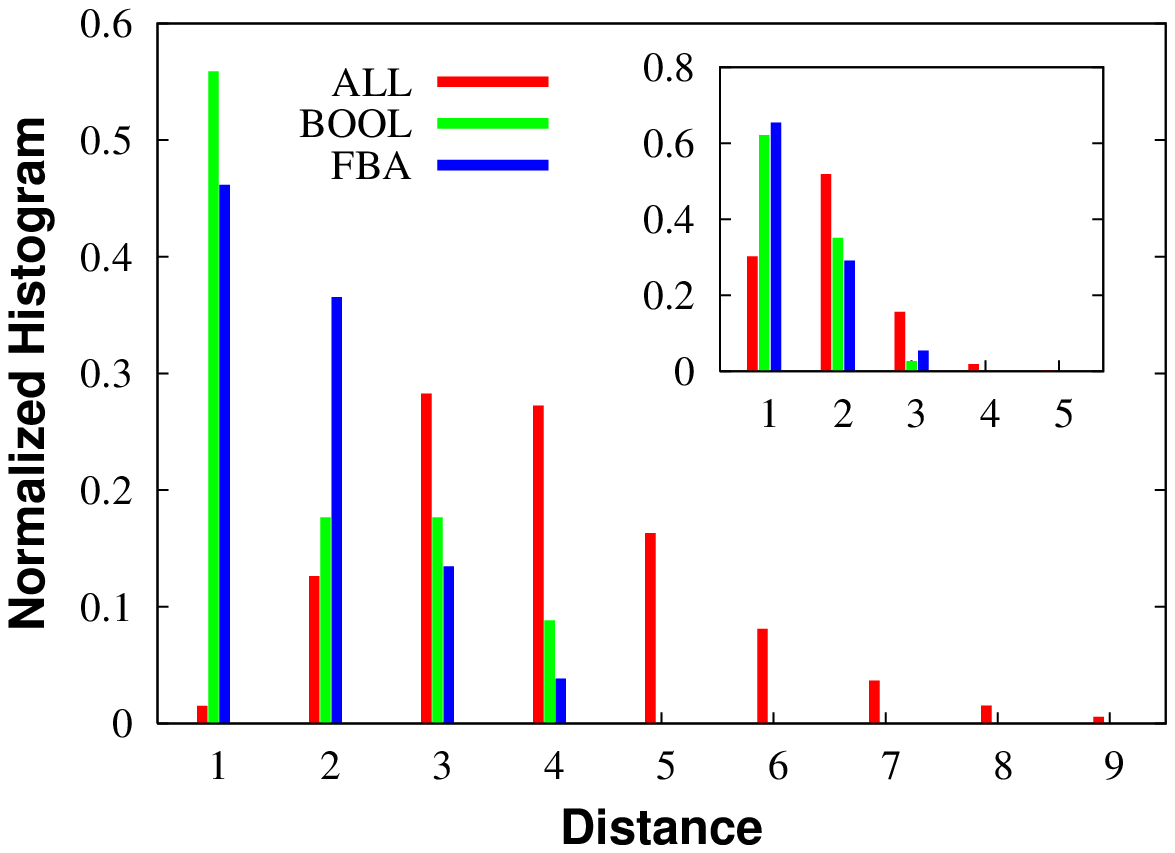}
\caption{Ghim, Goh \& Kahng}
\end{figure}

\newpage
\begin{figure}[h!]
\centering
\includegraphics[width=7cm,angle=-90]{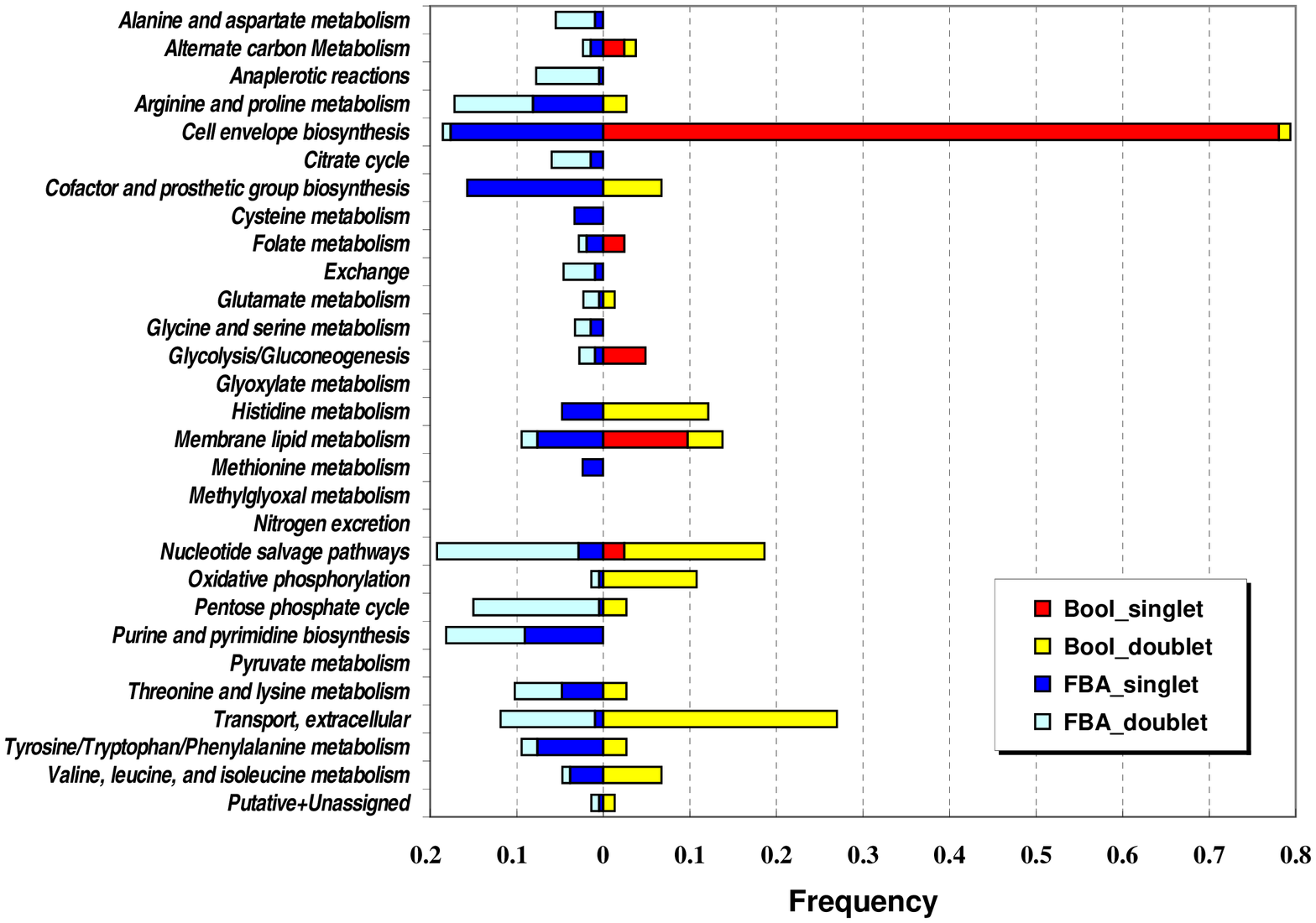}
\caption{Ghim, Goh \& Kahng}
\end{figure}

\begin{table}
\linespread 1
\tiny
\begin{center}
\begin{tabular}{llllcc}
\hline\hline
Reaction I & Reaction II & \hspace{20pt} Category I &\hspace{16pt} Category II & $\phi_{\textrm{I, wild}}$ & $\phi_{\textrm{II, wild}}$ \\
\hline
 ASNS2	& ASNS & Alanine, aspartate metabolism	&\hspace{30pt}\textquotedbl& 0.298933	& 0.000000\\
 ALAR	& ALARi	& Alanine, aspartate metabolism	&\hspace{30pt}\textquotedbl& 0.072057	& 0.000000\\
 PPC	& MALS	& Anaplerotic reactions	&\hspace{30pt}\textquotedbl& 3.519873	& 0.000000\\
 PPC	& ICL	& Anaplerotic reactions	&\hspace{30pt}\textquotedbl& 3.519873	& 0.000000\\
 DKMPPD2	& DKMPPD	& Arginine, Proline Metabolism	&\hspace{30pt}\textquotedbl& 0.009138	& 0.000000\\
 ORNDC	& ARGDC	& Arginine, Proline Metabolism	&\hspace{30pt}\textquotedbl& 0.054826	& 0.000000\\
 ORNDC	& AGMT	& Arginine, Proline Metabolism	&\hspace{30pt}\textquotedbl& 0.054826	& 0.000000\\
 GLUDy	& GLUSy	& Glutamate metabolism	&\hspace{30pt}\textquotedbl& -10.86474	& 0.000000\\
 KAS15	& KAS14	& Membrane Lipid Metabolism	&\hspace{30pt}\textquotedbl& 0.41942 & 0.000000\\
 ADK & ADK3	& Nucleotide Salvage Pathways	&\hspace{30pt}\textquotedbl& 3.284564	& 0.000000\\
 RNTR2	& RNDR2	& Nucleotide Salvage Pathways	&\hspace{30pt}\textquotedbl& 0.033157	& 0.000000\\
 RNTR2	& NDPK5	& Nucleotide Salvage Pathways	&\hspace{30pt}\textquotedbl& 0.033157	& 0.000000\\
 RNTR & NDPK8	& Nucleotide Salvage Pathways	&\hspace{30pt}\textquotedbl& 0.032243	& 0.000000\\
 RNDR3	& RNTR3	& Nucleotide Salvage Pathways	&\hspace{30pt}\textquotedbl& 0.033157	& 0.000000\\
 NDPK7	& RNTR3	& Nucleotide Salvage Pathways	&\hspace{30pt}\textquotedbl& 0.033157	& 0.000000\\
 NDPK & ADK3	& Nucleotide Salvage Pathways	&\hspace{30pt}\textquotedbl& 1.045572	& 0.000000\\
 GARFT	& GART	& Purine, Pyrimidine Biosynthesis	&\hspace{30pt}\textquotedbl& 0.623856	& 0.000000\\
 DHORD5	& DHORD2	& Purine, Pyrimidine Biosynthesis	&\hspace{30pt}\textquotedbl& 0.411327	& 0.000000\\
 O2t	& SUCCt2b	& Transport, Extracellular	&\hspace{30pt}\textquotedbl& 20.00000 & 0.000000\\
 PIt2r	& PIabc	& Transport, Extracellular	&\hspace{30pt}\textquotedbl& 1.190038	& 0.000000\\
 TRPS3	& TRPS & Tyrosine, Tryptophan, Phenylalanine Metabolism	&\hspace{30pt}\textquotedbl& 0.070490 & 0.000000\\
\hline
 RNDR & RNTR & Nucleotide Salvage Pathways	&\hspace{30pt}\textquotedbl& 0.269780 & 0.032243\\
 ADK & NDPK & Nucleotide Salvage Pathways	&\hspace{30pt}\textquotedbl& 3.284564	& 1.045572\\
 RPE	& TKT2	& Pentose Phosphate Cycle	&\hspace{30pt}\textquotedbl& 5.716916	& 2.573754\\
 RPE	& TKT & Pentose Phosphate Cycle	&\hspace{30pt}\textquotedbl& 5.716916	& 3.143163\\
 RPE	& TALA	& Pentose Phosphate Cycle	&\hspace{30pt}\textquotedbl& 5.716916	& 3.110267\\
 TALA	& TKT2	& Pentose Phosphate Cycle	&\hspace{30pt}\textquotedbl& 3.110267	& 2.573754\\
 TALA	& TKT & Pentose Phosphate Cycle	&\hspace{30pt}\textquotedbl& 3.110267	& 3.143163\\
 TKT & TKT2	& Pentose Phosphate Cycle	&\hspace{30pt}\textquotedbl& 3.143163	& 2.573754\\
 GND	& TKT2	& Pentose Phosphate Cycle	&\hspace{30pt}\textquotedbl& 10.42057	& 2.573754\\
 GND	& RPE	& Pentose Phosphate Cycle	&\hspace{30pt}\textquotedbl& 10.42057	& 5.716916\\
 THRAr	& THRS	& Threonine, Lysine Metabolism	&\hspace{30pt}\textquotedbl& -0.26978 & 0.405103\\
 THRAr	& HSK	& Threonine, Lysine Metabolism	&\hspace{30pt}\textquotedbl& -0.26978 & 0.405103\\
\hline\hline
 ORNDC	& UREAt	& Arginine, Proline Metabolism	& Transport, Extracellular	& 0.054826	& 0.000000\\
 ORNDC	& EX\_urea	& Arginine, Proline Metabolism	& Exchange & 0.054826	& 0.000000\\
 GALUi	& GALU	& Cell Envelope Biosynthesis	& Alternate Carbon Metabolism	& 0.025847	& 0.000000\\
 FUM	& SUCCt2b	& Citrate Cycle (TCA)	& Transport, Extracellular	& 1.368573	& 0.000000\\
 FRD2	& DHORD2	& Citrate Cycle (TCA)	& Purine, Pyrimidine Biosynthesis	& 0.411327	& 0.000000\\
 MTHFD	& GART	& Folate Metabolism	& Purine, Pyrimidine Biosynthesis	& 1.365196	& 0.000000\\
 MTHFC	& GART	& Folate Metabolism	& Purine, Pyrimidine Biosynthesis	& 1.365196	& 0.000000\\
 CBMK	& CBPS	& Putative	& Arginine, Proline Metabolism	& 0.77814 & 0.000000\\
 VALTA	& VPAMT	& Valine, leucine, isoleucine metabolism	& Alanine, aspartate metabolism	& -0.524765	& 0.000000\\
\hline
 SUCD1i	& PPC	& Citrate Cycle (TCA)	& Anaplerotic reactions	&0.414887	& 3.519873\\
 FUM	& PPC	& Citrate Cycle (TCA)	& Anaplerotic reactions	&1.368573	& 3.519873\\
 SUCD4	& PPC	& Oxidative phosphorylation	& Anaplerotic reactions	&0.414887	&3.519873\\
 GARFT	& MTHFD	& Purine, Pyrimidine Biosynthesis& Folate Metabolism	&0.623856	&1.365196\\
 GARFT	& MTHFC	& Purine, Pyrimidine Biosynthesis& Folate Metabolism	&0.623856	&1.365196\\
 THRS	& GHMT2	& Threonine, Lysine Metabolism	& Glycine, Serine Metabolism	&0.405103&1.653370\\
 HSK	& GHMT2	& Threonine, Lysine Metabolism	& Glycine, Serine Metabolism	&0.405103&1.653370\\
 O2t	& PPC	& Transport, Extracellular	& Anaplerotic reactions	&20.00000	&3.519873\\
 O2t	& DKMPPD2& Transport, Extracellular	& Arginine, Proline Metabolism	&20.00000&0.009138\\
 O2t	& FRD2	& Transport, Extracellular	& Citrate Cycle (TCA)	&20.00000	&0.411327\\
 O2t	& GAPD	& Transport, Extracellular	& Glycolysis/Gluconeogenesis	&20.00000&32.61301\\
 O2t	& PGK	& Transport, Extracellular	& Glycolysis/Gluconeogenesis	&20.00000&-32.61301\\
 O2t	& DHORD5& Transport, Extracellular	& Purine, Pyrimidine Biosynthesis&20.00000	&0.411327\\
\hline\hline
\end{tabular}
\end{center}
\caption{\small Ghim, Goh \& Kahng}
\end{table}

\end{document}